\documentclass{elsart}
\usepackage{latexsym, graphicx}

\def\gamp{{\tt gamp }}

\begin{document}

\date{9 September 2003}

\bibliographystyle{elsart-num}

\begin{frontmatter}
\title{An Object-Oriented Approach to Partial Wave Analysis}
\author{John P. Cummings\corauthref{cor}}
\address{Rensselaer Polytechnic Institute}
\ead{cummij@rpi.edu}
\author{Dennis P. Weygand}
\address{Jefferson Laboratory}
\ead{weygand@jlab.org}
\corauth[cor]{Corresponding author. Address: Department of Physics, Rensselaer Polytechnic Institute, 110 8th Street, Troy, NY, 12180-3590 Phone: (518) 276-2542}

\begin{abstract}
Partial Wave Analysis has traditionally been carried out using a set
of tools handcrafted for each experiment.  By taking an object-oriented
approach,  the design presented in this paper attempts to create a more
generally useful, and easily extensible, environment for analyzing
many different type of data.  
\end{abstract}

\end{frontmatter}

\section{Introduction}

Partial Wave Analysis, or PWA, is a technique used in hadron spectroscopy
to extract information about the spin-parity and decay properties
of resonances produced in hadronic interactions.  Typically, these
resonances are produced at accelerator experiments via a variety of
production mechanisms.  The resonances produced by these methods also
appear in the decay products of other well known resonances, such as
the $J/\psi$, and can be studied there.  Although the tools we describe
here could be used in an investigation that uses any of these production
mechanisms, perhaps with slight modification,  they have been
used extensively only in peripheral production experiments, and so we will use this
type of production as an illustrative example in this paper.  Currently, new results are being obtained studying baryon resonances produced in $s$-channel $ \gamma p $ interactions.

The general idea is to parameterize the intensity distribution in terms
of variables that have physical meaning when interpreted as properties of 
intermediate states in a particular reaction.  In principle, any
complete set of functions which span the appropriate space can be used.
Although many parameterizations are possible, for instance an (almost)
purely mathematical description in terms of the moments, we choose an
expansion in terms of intermediate resonances and their decays.  This has
at least two advantages. Firstly, it allows us to take advantage of
physics such as conservation laws to limit the number of terms we must
include in our expansion to get a good description of the intensity
distribution.  Secondly, it allows a more direct interpretation of
our results.  A moment analysis, for instance, requires a complicated
mapping from moments to physical states to understand the results in
all but the simplest of cases.

The formalism we have used in implementing our system is
based on the papers of Chung~\cite{th:chung88}
and Chung and Trueman.~\cite{th:chung75}
The intensity distribution is written
as a sum of amplitudes, squared appropriately to account for interference:
\begin{equation}
I(\tau)~=~\sum_\alpha \left\{  \left|  \sum_\beta
{}^\alpha \psi_\beta \left(\tau \right) \right|^2 \right\}.
\end{equation}
The variable $\tau$ represents the set of variables necessary to define
a configuration of the final state being investigated.  It typically
includes the angles of the decay products in various reference frames,
masses of two body sub-systems, {\em etc}.  The subscripts $ \alpha $ and $ \beta$ are
the parameters that describe the partial wave decomposition we are using,
$ \alpha$ specifying properties of the different intermediate states that
do not interfere, such as the spin states of the incoming or outgoing
particles in the detector.  The subscript $ \beta$, on the other hand,
represents the properties whose differing values do interfere, for
instance the spin states of broad resonances produced as intermediate
states in a sequential decay.

\begin{figure}
\begin{center}
\includegraphics{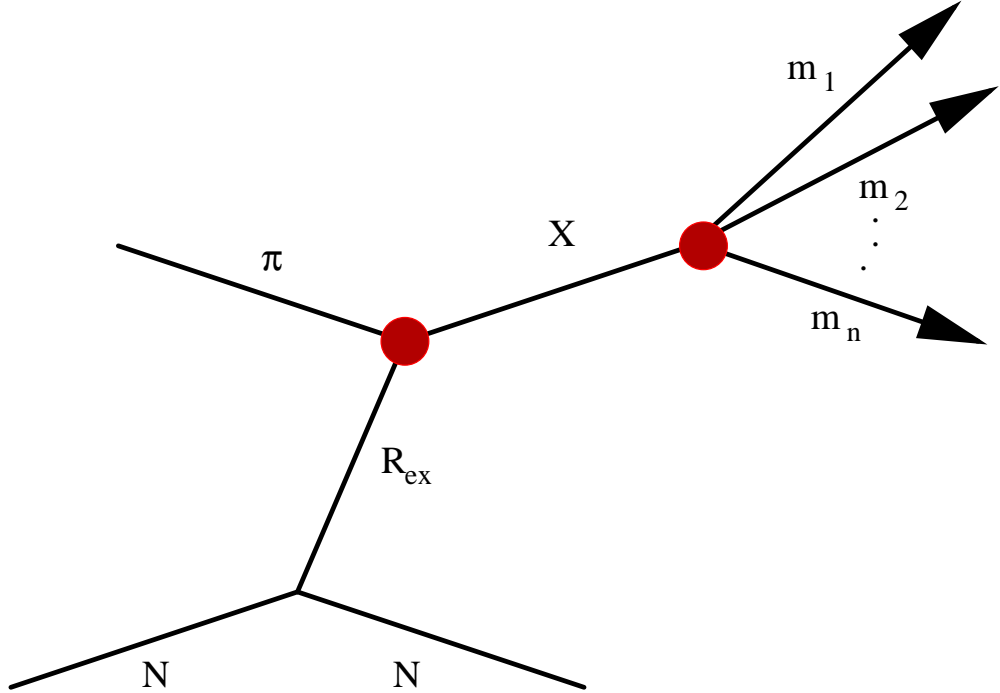}

\caption{\label{expansion_diagram}Diagram representing the amplitude $ \psi_{ \alpha \beta } \left( \tau \right)$.  The intermediate state $X$ is produced by the exchange of a Reggion $R_{ex}$ between the $ \pi$ beam and the nucleon target.}
\end{center}
\end{figure}

In a peripheral production experiment, the amplitudes in the expansion
can be drawn as shown in Fig.~\ref{expansion_diagram}.
Guided by this picture, the amplitude
$ \psi_{ \alpha \beta } \left( \tau \right) $ is factored into two parts: $V$,
the amplitude to produce the state $X$, and $A$, the amplitude for the
state $X$ to decay into the final state observed.

These amplitudes are written in the reflectivity
basis~\cite{th:chung75}, which takes into account parity conservation in
the production process by writing the amplitudes in terms of eigenstates
of reflection through the production plane.  The reflectivity of the
amplitude, denoted $\epsilon$, is defined so that in the case of a pion
beam it coincides with the naturality of the exchanged Regge trajectory.
Waves of differing $\epsilon$ do not interfere.  Also, amplitudes with
different relative spin configurations for the incoming and outgoing
baryon will not interfere.  The spin configuration of the amplitudes is
labeled by $k$, and the number of allowed values, sometimes referred to
as the rank of the fit, is typically 2 for a spin $\half$ recoil baryon.

In the reflectivity basis our sum over amplitudes splits into four
non-interfering sets of fully interfering amplitudes, or $\alpha~=~\left\{ \epsilon , k \right \}$.  While both the production and decay
amplitudes depend on $\epsilon$, the decay amplitude does not depend on
$k$, since the decay amplitude for a particular state $X$ cannot depend
on what the proton spin did during the production process.  Similarly,
the production amplitude does not depend on $\tau$, the configuration
of the final state the $X$ decays into.

The intensity distribution now becomes
\begin{equation}
I \left( \tau \right)~=~\sum_{ \epsilon , k } \left\{
\left| \sum_\beta {}^\epsilon  V_{ k \beta }
{}^\epsilon  A_\beta \left( \tau \right) \right|^2 \right\}.
\end{equation}
The decay amplitudes $ {}^\epsilon A_{ k \beta } \left( \tau \right)$
can be calculated for each event.  By varying the unknown production
amplitudes ${}^\epsilon V_{ k \beta }$ the predicted intensity
distribution above is matched as closely as possible by the observed
intensity as a function of all kinematic variables.  This is done through
an extended maximum likelihood fit.  During the fitting process, the
finite acceptance of the detector is taken into account on a term by term
basis, {\em i.e.}, each term contains a pair of of decay amplitudes with a unique shape in $\tau$, and the acceptance for each term is determined separately.

The likelihood function is defined as a product of probabilities,
\begin{equation}
\mathcal{L} = \left[ {\bar{n}^n\over{n!}} e^{-\bar{n}}\right]\prod_i^n \left[{I(\tau_i)}\over{\int(\tau)\eta(\tau)d\tau}\right]
\end{equation}
The term outside the product is the Poisson probability of observing $n$
events and reflects the fact that we use the extended maximum likelihood
method.  The integral in the denominator of the summed term contains the
acceptance $\eta(\tau)$ and is referred to as the accepted normalization
integral.

The function which is actually maximized in the likelihood fit is
\begin{equation}
\ln { \mathcal{L}} = \sum_i^n \left[ \ln\sum_{ k, \epsilon , \beta , \beta^\prime } {}^\epsilon 
V_{ k \beta } {}^\epsilon V_{ k \beta^\prime }^{*} {}^\epsilon 
A_{ \beta }( \tau_i ) {}^\epsilon A_{ \beta^\prime }^{*} ( \tau_i ) \right] -
n \left[ \sum_{ k, \epsilon , \beta , \beta^\prime } {}^\epsilon V_{k \beta } {}^\epsilon 
V_{k \beta^\prime }^{*} {}^\epsilon \Psi_{ \beta \beta^\prime }^x \right]
\label{ln_like}
\end{equation}
The first sum is over data events, where the term being summed over is
simply the intensity $I( \tau_i )$ for each event.  The second term contains
the accepted normalization integrals ${}^\epsilon \Psi^x_{ \beta \beta^\prime }$
where the superscript $x$ denotes accepted.  This integral is evaluated
numerically:
\begin{equation}
{}^\epsilon \Psi^x_{ \beta \beta^\prime } = { 1 \over M_x } \sum_i^{ M_x } {}^\epsilon 
A_{ \beta } ( \tau_i ) {}^\epsilon A^{*}_{ \beta^\prime }( \tau_i ) .
\label{norm_int}
\end{equation}
The sum is over an accepted Monte-Carlo data set of $M_x$ events.
A similar integral is calculated for the raw Monte-Carlo data set, to
be used in the calculation of observables.  For instance, the number of
acceptance corrected events the fit predicts is
\begin{equation}
N = { n \over { \eta_x }} \sum_{ \epsilon k \beta \beta^\prime } {}^\epsilon 
V_{ k \beta } {}^\epsilon V^{*}_{ k \beta^\prime } {}^\epsilon \Psi_{ \beta \beta^\prime },
\label{eq:nevents}
\end{equation}
where $ {}^\epsilon \Psi_{ \beta \beta^\prime }$ is the raw normalization integral.
By varying the range of the values of $\left\{ \epsilon k \beta \beta^\prime\right\}$
included   in the sum, the number of events due to different combinations
of amplitudes can be determined.

\subsection{Decay Amplitudes}

Calculation of decay amplitudes for the resonance $X$ is done
recursively using the isobar model, regarding the $n$-body final state as
the result of a series of sequential decays (usually two body) through
intermediate states known as the isobars.  The amplitude for $X$ to decay
into the final state is then simply the amplitude for $X$ to decay into
its immediate children times the amplitude for each of its children to
decay.

\begin{figure}
\begin{center}
\includegraphics{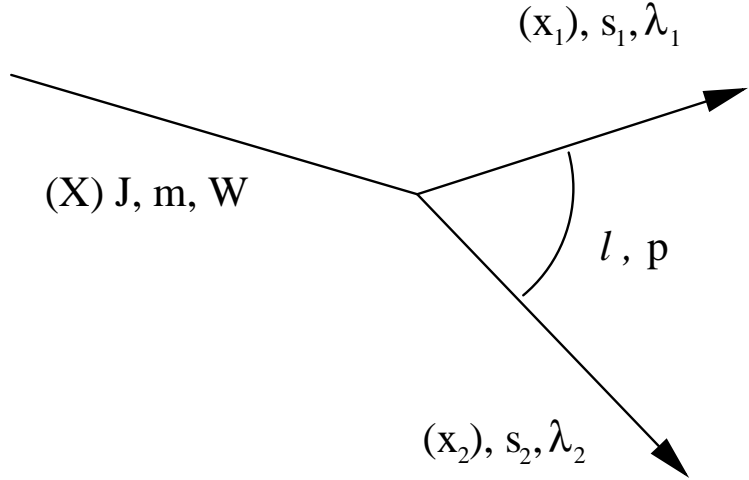}
\caption{\label{GJdecay}Decay of the resonance $X$ into $x_1$ and $x_2$.}
\end{center}
\end{figure}

The initial decay of the $X$ into its children is evaluated in the
Gottfried-Jackson frame.  This frame is a rest frame of the resonance
$X$ with the $z$ axis in the direction of the beam and the $y$ axis
perpendicular to the production plane. The quantities used in defining
the amplitude are shown schematically in Fig.~\ref{GJdecay}

The state $X$ of mass $W$ has spin $J$ with $z$ projection $m$.  It
decays into children $x_i$ with spin $\sigma_i$ and helicity
$\lambda_i$. These children have a breakup momentum $\vec{p}$ and
relative orbital angular momentum $\ell$.  The decay amplitude can then be
written down~\cite{th:chung71}

\begin{equation}
A_X = \tilde{\ell} \sum_\lambda D^{J *}_{ m \lambda } ( \Omega )( \ell 0 s \lambda | J \lambda )( s_1 \lambda_1 s_2 -\lambda_2 | s \lambda ) F_\ell ( p ) a_{ \ell s } A_{x_1} A_{x_2}
\end{equation}

where $\lambda = \lambda_1 - \lambda_2$ and $ \vec{s}  = \vec{s}_1 + \vec{s}_2$, 
{\em i.e.} , $\vec{s}$ is the total spin of the two children and
$\lambda$ is the component of $\vec{s}$ in the direction defined by
$x_i$'s momentum.  The $A_{x_i}$ are the decay amplitudes of each child.

The $\tilde{\ell}  = ( 2 \ell + 1 )^{ \half }$ factor, along with the two
Clebsch-Gordon coefficients, come from the fact that we are using
helicity states and must relate the helicity coupling constant to the
$\ell s$-coupling constant, $a_{\ell s}$, we want to find through the partial
wave expansion.\cite{th:chung71}

Rotational properties of the helicity states lead to the introduction of
$D$-function $D^{ J * }_{ m \lambda } ( \Omega )$ where $\Omega = ( \theta
, \phi ,0)$ are the Euler angles of $x_1$ in the Gottfried-Jackson frame.
The choice of the third angle $\gamma$ defines the phase convention.
We choose $\gamma = 0$, which is different from  that of Jacob and
Wick~\cite{Jacob:1959at}, who choose $ \gamma = -\phi$.

$F_\ell ( p )$ is an angular momentum barrier factor added to give the
amplitude the correct behavior near threshold.   We used the
Blatt-Weisskopf centrifugal-barrier functions as given by von Hippel
and Quigg.~\cite{th:vonhippel72}

Finally, $a_{\ell s}$ is the $\ell s$ coupling constant which
contains the dynamics of the decay.  This factor is absorbed into the
production amplitude for this wave, which is then determined through
the fit.  The coupling constant $a_{\ell s}$ is in general a function
of the mass $W$ of the state $X$, and this is usually handled in the
fits by performing them in bins of $W$ which are narrow enough to
assume the $a_{\ell s}$ constant over the width of the bin.

\begin{figure}
\begin{center}
\includegraphics{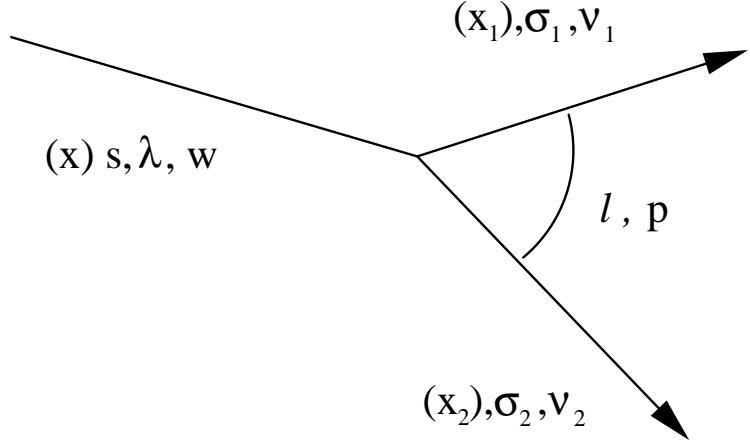}
\caption{\label{Hdecay}Decay of a child $x$ into $x_1$ and $x_2$.}
\end{center}
\end{figure}

The decay amplitudes of the children $A_{x_i}$, and recursively their
children, {\em etc.}, 
are calculated in the helicity frame of the
child that is decaying.  For instance, consider one of the $x_i$ above,
but now we call it simply $x$ since it has become the parent particle of a new decay, as
in Fig.~\ref{Hdecay}

Here the particle decaying, $x$ has mass $w$, spin $s$, and helicity
$\lambda$.  Its decay products $x_i$ have spins $\sigma_i$ and
helicities $\nu_i$, with relative orbital angular momentum $\ell$ and
breakup momentum $p$.

The form of the amplitude for this piece of the decay is very similar
to what was written above
\begin{equation}
A_x = \tilde{\ell} \sum_\lambda D^{s *}_{ \lambda \nu } ( \Omega )(\ell 0 \sigma \nu | s \nu )( \sigma_1 \nu_1 \sigma_2 - \nu_2 | s \nu ) F_\ell (p) \Delta_x (w) A_{x_1} A_{x_2}
\end{equation}
where again $\nu = \nu_1 - \nu_2$ and $\vec{\sigma} = \vec{\sigma}_1 + \vec{\sigma}_2$

The pieces of this decay that look similar to the initial decay in the
Gottfried--Jackson frame do so because they come from the same place.
The same Clebsch--Gordon coefficients and $\tilde{\ell}$ factor show up from the
relationship of the helicity to $\ell s$ states, the $D$--function from
the rotational properties of the states, and the angular momentum
barrier factor suppresses high-$\ell$ decays near threshold where the
breakup momentum is small.

The only difference, in fact, is the appearance of $\Delta_x(w)$ in
place of the coupling constant $a_{\ell s}$ from the initial decay.  In
the decays of the isobars it is usually assumed that the dynamics of
the decay, which depend on the isobar mass $w$, are known and put in
explicitly.  Often this is done as a simple relativistic Breit-Wigner,
although it is sometimes necessary to use a more complex
parameterization, such as coupled-channel Breit-Wigners~\cite{th:flatte76}
or a $K$--matrix parameterization~\cite{Chung:1995dx}.

Total decay amplitudes are made up as a product of these intermediate
decays.  In order to parameterize the spin-density matrix, which is
determined by the fit, in the simplest way, it turns out to be
advantageous to transform the states into the reflectivity basis.  The
reflectivity basis is defined by eigenstates of reflection in the
production plane,  the details of the transformation can be found in
Chung and Trueman.~\cite{th:chung75}



\section{Choice of Tools}

After determining the design criteria from the considerations in the
last section,  we choose tools to assist in building the system to meet
these criteria.

\subsection{The choice of language}

Choosing a language was not as difficult as one might hope: the paucity
of object-oriented languages having the support of adequate development
tools led us almost immediately to C++.  Java, Eiffel and Sather were
briefly considered.  Java was decided to still be too slow for numerically
intensive calculations.  Eiffel and it's relative Sather are interesting
languages that address some of the deficiencies of C++, but the lack
or weakness of development tools such as a symbolic debugger led us to
pass them up.

C++ is a popular language which now has an {\sc ANSI/ISO} standard.  There are
numerous compilers, debuggers, {\sc CASE} tools and class libraries available.
In particular, we choose the {\sc GNU} compiler suite and development tools.
The {\sc GNU} C++ compiler is very close to the {\sc ANSI} standard, and since most
of our development and analysis is done on a Linux/{\sc GNU} platform it is
also the native compiler on these systems.
Using C++ also gives us the advantages of an object-oriented language
while also providing convenient access to useful libraries written in
C or FORTRAN.  For instance the engine of our fitting program is the
CERNLIB package MINUIT.

Attempting to deliver a ``turnkey'' system that, given a physics
data set in a standard format, would allow users to perform an analysis
implies a relatively sophisticated control language for our tools.
Yacc and it's cousin Lex allowed easy development of a 
flexible grammar for driving both the decay amplitude calculator and
the fitter.  These development tools interface very well with C++ and
are standard on any *NIX system.
\begin{figure}
\begin{center}
\includegraphics[width=4in]{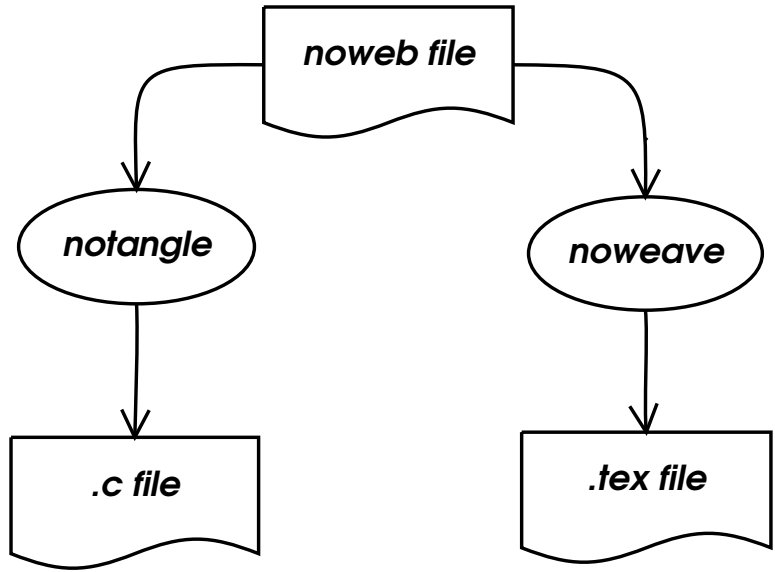}
\caption{\label{noweb}A {\tt noweb} source file is pre-processed with either {\tt notangle} or {\tt noweave} to produce either source code or documentation, respectively.}
\end{center}
\end{figure}

\subsection{The choice of documentation system}

In our experience, up to date documentation for software written by
physicist is hard to find--our own code has not been an exception.
We take the view that self-documenting code is a realizable goal
{\em as long as the audience is programmers}.
For this project we were striving to reach a much broader audience,
and hence recognized the need for something else.  We decided to use 
{\tt noweb}~\cite{comp:noweb}, 
a simple 
{\em literate programming}
package.

Literate programming~\cite{book:literateprogramming} was first proposed by Donald Knuth, which he
implemented in the form of
{\tt web}.
Essentially code and documentation are written interspersed in the
same files, making it easier to document the code as it is written,
and encourages the documentation to keep pace with program development.
The pre-processors which make up the
{\tt web}
system then
{\tt weave}
the web file into documentation, usually in latex format, or
{\tt tangle}
the web file into source code.  While some literate programming tools
are language specific,
{\tt noweb}
allows any programming language to be embedded into its files, allowing
symmetric treatment of C++, yacc, lex,
{\em etc}.

\section{The Design}

\subsection{The architecture of the suite}

The design of the tools is strongly influenced by the mathematics of the
fitting, some of which was described above.  In general, we felt that a 
set of independent programs each of which which work with the output of the others is
the best design.  We clearly need a program to perform the minimization
of the $- \ln ( \mathcal{L} )$ function of equation~\ref{ln_like}.
The normalization integral, equation~\ref{norm_int}, is calculated prior to
fit and requires a program of its own.  
We must have a program to calculate decay amplitudes as a function of
$\tau$.  This implies three basic programs are required for an analysis.
During the course of testing and use, however, several additional utility
programs were written.  Some of these turned out to be generally useful
and will also be briefly described in this report.

\begin{figure}
\includegraphics[width=5in]{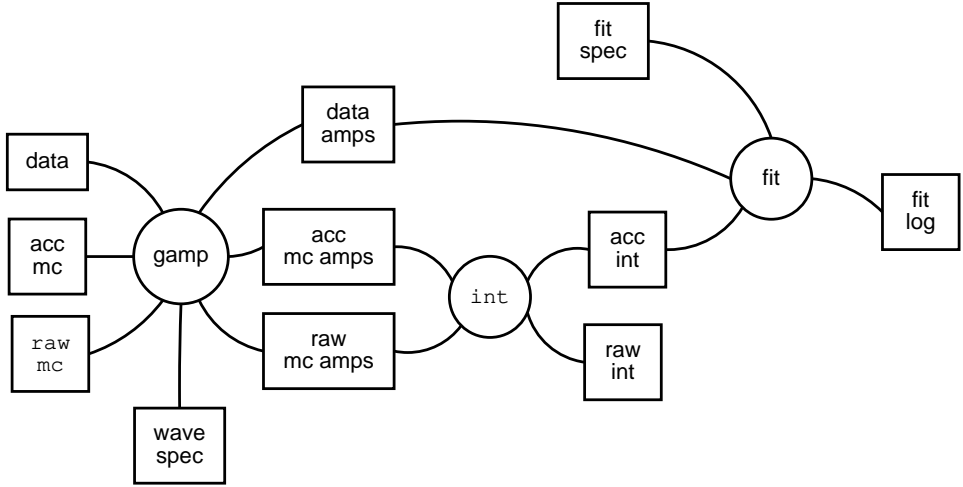}
\caption{\label{pwa_flow}Pictorial view of the entire PWA procedure.  The three circles represent the three major computational steps, acting on the rectangular data files. }
\end{figure}

A picture of the entire procedure is shown in Fig.~\ref{pwa_flow}.
The decay amplitude calculator {\tt gamp} computes the decay amplitude
given an event and a particular ``wave'': an assumed spin-parity for
the whole system, any isobars with their spin-parities, and a orbital
angular momentum and total spin for each two body decay.  The integrator
{\tt int}, used to compute normalization integrals used in the fitting
process, must read amplitudes calculated by {\tt gamp} and integrate
over the appropriate space.  The production amplitudes are found by {\tt
fit} using a maximum likelihood fit.

\subsection{The architecture of gamp}

A typical invocation of the program to generate amplitudes might look like:
\begin{quotation}
{ \tt zcat data.gamp.gz | gamp 1++0+rho\_pi.key > 1++0+rho\_pi.amps }
\end{quotation}
Notice that
{\tt gamp}
reads and writes from standard input and output, and that it's single
required argument is the name of a
{\em  keyfile},
a program specifying the amplitude to be calculated for each event read
from it's input.
\gamp
uses an
{\tt awk}-like
processing model: the program in the keyfile is run repeatedly on each
event as it is read in.  The language of this program is described 
below.

\subsubsection{Event format}

The format of the events read from standard input is quite simple. 
The events are expected to be in {\sc ASCII} format. Each event begins with
an integer specifying the number of particles which will follow.
Particles are given by their {\sc GEANT} id (integer), their charge (integer), 
and the four components of their four momentum, $p_x$, $p_y$, 
$p_z$, $E$ (floats). 
The first particle in each event is always the beam.  Presently, the
target is not specified and is assumed to be a proton (this is likely
to change in the near future, and the target will be required).  
Following the beam come all final state particles from the system being
analyzed, in arbitrary order, and then the next event,
{\em etc}.

For example, the top of a file containing events from the reaction
$ \pi^- p \rightarrow p \eta \pi^+ \pi^- \pi^- $ might look like
\begin{verbatim}
5
9 -1  0.0003765 -0.0573188 18.3642158 18.3648357
9 -1  0.0814732  0.4102049  2.2733631  2.3157212
9 -1  0.2404482  0.3578656  3.9542443  3.9801270
8  1  0.3348054  0.0241324  3.3473827  3.3670651
17 0 -0.4792622 -0.4030199  8.5089292  8.5494852
5
9 -1 -0.0850012 -0.0702078 18.1182842 18.1191577
9 -1 -0.4165566  0.1685560  1.5984030  1.6662241
...
\end{verbatim}

Notice in this example, from BNL E852, the reaction is assumed to be
a $t$-channel process.  Therefore the system being studied here is 
$\eta \pi \pi \pi$ and the final state proton does not appear in the input.  This is
optional, the keyfile specifying the amplitude to calculate will refer
to particles by name, unmentioned particles are simply ignored.  Thus
specifying the proton in the event and ignoring it in the keyfile will
not create an error. 
If an $s$-channel amplitude is being calculated,
{\em all}
final state particles would be required in the event specification.

Since
{\tt gamp}
reads events from standard input, these ASCII files can be stored compressed
and then
{\tt zcat}ted
into a pipe.  Or better still, never put to disk, but created on the fly 
from the (presumably) more efficient binary data format of the particular
experiment.

\subsubsection{The keyfile language}

The keyfile language is fairly simple.  
The file is free-format:
spaces, tabs, and newlines may be used to improve readability.
Statements are terminated by semi-colons.

A very simple keyfile might look like this:

\begin{verbatim}
# debug = 1;
channel = t;
mode = binary;

J = 2 P = +1 M = 2 {
        pi+
        pi-
        l=4
}
;
\end{verbatim}

In this file we are specifying the decay of a state with $J^P = 1^+$
and $m = 1$ into two pions with two units of angular momentum between
them, {\em i.e.}, an D wave.

A {\tt  \#} comments the remainder of the line.  Here the {\tt debug =
1;}, which would turn on copious debugging information, is commented out.
The amplitude being calculated is for a $t$-channel process.  The only
allowed values for {\tt channel } are {\tt s} and {\tt t}.  Amplitudes
output can be in either {\sc ASCII} (default) or binary format, and can be
selected by setting {\tt  mode} to either {\tt ascii} or {\tt binary}.
Finally comes the specification of the wave itself. A wave is a set of
quantum numbers--$J$, $P$ and $M$--followed by a decay.  {\bf Note}:
all angular momenta are in units of $\hbar/2$, {\em  i.e.}, multiplied
by 2.  This allows particles with half-integer spin to be represented
as integers in the input.  In the example above, for instance, the state
being calculated has $J^P = 1^+$.

The decay is represented by a pair of child particles followed by the
orbital angular momentum between them, and optionally, their total spin.
If the total spin is not ambiguous, it may be omitted.  Only when both
particles have non-zero spin is it necessary to explicitly give the spin,
and it {\em must} follow the orbital angular momentum.  The orbital
angular momentum and spin may be labeled as above or may be unadorned
integers.  The first integer is always the orbital angular momentum
and the second, if present is the total spin.  The decay specification
is delimited by the pair of curly brackets.
Although it may be spread out over many lines, the entire wave is a
single statement, so don't forget the {\tt ;} at the end.

\subsubsection{Combining Waves}

Waves may be added to eachother, and this gives us a way to handle
symmetrizing amplitudes when there are identical particles. If the final
state contains identical particles, they are identified in the wave
specification as if the identical particles were read into a (1-indexed)
array, {{\em i.e.}, {\tt pi+[1]} is the first positively charged pion
read for each event, {\tt pi+[2]} is the second {\em etc}.  So to form
a Bose-symmetrized amplitude use something similar to the following:

\begin{verbatim}
# debug = 1;
channel = t;
mode = binary;

0.707 * (
J = 4 P = +1 M = 0 {
        pi0[1]
        pi0[2]
        l=4
}
+ J = 4 P = +1 M = 0 {
        pi0[2]
        pi0[1]
        l=4
} )
;
\end{verbatim}

In this case the second wave look the same with the exception of the
flipped indices on the {\tt pi0}'s: we have exchanged the two identical
particles.

\subsubsection{Sequential Decays}

Any unstable particle may be followed by a similar decay specification,
recursively.  For example, {\tt  gamp} would not balk at a decay chain
as follows:

\begin{verbatim}
# debug = 1;
channel = t;
mode = binary;

J = 2 P = +1 M = 0 {
        a2(1320) {
                rho(770) {
                        pi+[1]
                        pi-[1]
                        l = 2
                }
                pi0
                l = 4
        }
        rho(770) {
                pi+[2]
                pi-[2]
                l = 2
        }
        l = 2
        s = 2
} ;
\end{verbatim}

This keyfile describes a $J^P=1^+$ particle in an $m=0$ state decaying into $a_2 \rho$ via a $P$ wave.
Notice here that we have included a specification of the total spin of
the $a_2 \rho$ system $s=1$, as it is not unique.  The $a_2(1320)$ decays further in this example into $\rho \pi$ in a $D$ wave, and finally each $\rho$ decays into $\pi \pi$ with $\ell = 1$.

\subsubsection{Mass Dependencies}

Currently, {\tt  gamp} also understands a few different mass dependencies
for isobars: flat, Breit-Wigner, and two solutions for $( \pi \pi )_S$.
Others will be added by popular demand.  The alternate mass dependencies
are given by appending, for instance, the statement {\tt massdep = flat}
after the decay, as in
\begin{verbatim}
# debug = 1;
channel = t;
mode = binary;

0.707 * (
J = 4 P = +1 M = 0 {
        pi0[1]
        pi0[2]
        l=4
} massdep=flat
+ J = 4 P = +1 M = 0 {
        pi0[2]
        pi0[1]
        l=4
} massdep=flat )
;
\end{verbatim}
The other appropriate keywords are {\tt  bw} for Breit-Wigner
(the default), {\tt  amp} for Au-Morgan-Pennington $( \pi \pi )_S$
parameterization (M-solution)~\cite{th:amp}, and {\tt  amp\_ves} for VES
modification~\cite{ex:amp_ves,Chung:2002pu} to above.

\subsubsection{Helicity Sums}

The default action when {\tt  gamp} sees a particle with spin is to sum
over the allowed helicities.  This is in general the correct action as
particles with spin are usually intermediate states which interfere and
must be summed over at the amplitude level.  Final state particles are
typically spinless with the notable exceptions of protons and neutrons.
If these appear in the wave specification the amplitude must {\em not}
be summed over and the {\tt helicity} or {\tt h} keyword allows for this.

\begin{verbatim}
mode=binary;
channel=s;

0.94868 * (
J=3 P=-1 M=1 {
        delta(1232)[1] {
                p+[1]   h=-1
                pi+[1]
                2
        }
        pi-[1]
        4
}
+ 0.33333 * (
J=3 P=-1 M=1 {
        delta(1232)[1] {
                p+[1]   h=-1
                pi-[1]
                2
        }
        pi+[1]
        4
}
));
\end{verbatim}

This keyfile describes the decay of a $J^P={3\over{2}}^-$ baryon resonance
into $\Delta \pi$ in a $D$ wave.  The two charge possibilities for the
$\Delta$'s are combined with the correct Clebsch-Gordon coefficients to
produce an isospin $1\over 2$ state.  This wave is being calculated for
the negative helicity of the final state proton; presumably the positive
helicity would be calculated also and added incoherently at fit time.

\subsection{The architecture of int}

The normalization integrals, whose values are needed at the time of the
fit, are calculated using {\tt int}.  Recall that the integrals needed
look like
\begin{equation}
\Psi_{ \alpha \alpha^\prime } = \int \psi_\alpha^* ( \tau ) \psi_{ \alpha^\prime } ( \tau ) \eta ( \tau ) d \tau ~=~ {1 \over M} \sum_i \psi_\alpha^* ( \tau_i ) \psi_{ \alpha^\prime } ( \tau_i )
\end{equation}
where $d \tau$ is an element of phase space, $\tau$ is then a point in
phase space, $\psi_\alpha ( \tau )$ is the decay amplitude for the wave $
\alpha $ as a function of the kinematic variables defining phase space
$\tau$.  This is an accepted integral, used at fitting time to perform
the acceptance correction, and hence the appearance of the acceptance
$\eta ( \tau ) $.  Unfortunately,  the acceptance of a detector is rarely
known analytically, and
so this integral is always done numerically. This is shown in the last
term  as a sum over Monte-Carlo generated events.  The generation is done
uniformly in phase space, {\em i. e.} flat in $ \tau $.  These events
are then passed through a detector simulation program,
and subjected to the same analysis and cuts.  Using only these $M$
remaining events in the sum above take into account the $\eta ( \tau )
$ factor in the integral.  Similar integrals are needed {\em post}-fit
to calculate observables from the fit results, as descibed by equation~\ref{eq:nevents}.  These ``raw'' integrals
differ from the above ``accepted'' integrals only by their lack of the
acceptance term $\eta ( \tau ) $ in the raw integrals.  Numerically this
corresponds to performing the sum over the entire generated phase space
Monte-Carlo event sample.

{\tt int} assumes all decay amplitudes $\psi_\alpha ( \tau_i )$ are
available on disk and a single file holds all amplitudes for a single wave
$\alpha$ for each event in a particular data set.  All the amplitude files
should correspond to the same data set.  The integrals $\Psi_{ \alpha
\alpha^\prime }$ are kept internally as matrices.  Individual integrals
may be accessed by either integer indices of the matrix, or by a string
representing the name of the wave--a human readable form of $\alpha$.

\subsection{The architecture of fit}

In the course of an analysis many fits need to be done.  It is important
to try different sets of waves in many combinations: two waves may not
be important individually, but their interference may be.  In addition,
once the best set of waves is determined, the stability of the fit can
be studied by repeating the fit with varying starting values for the
parameters.   A completed analysis will often require hundreds of fits.

The likelihood function maximized varies depending on the type of
process being modeled.  A $s$-channel likelihood function differs from
a $t$-channel function, and a photon beam requires a different function
from a pion beam.  Different assumptions may also be tested with regard
to the ``rank'' of the fit, relating to whether or not the amplitudes
have differing dependence on spin degrees of freedom.

The lack of a standard function led to the idea of using an interpreter
for the likelihood function.  The input file for the fit contains
the specification of the likelihood function.  A simple example file
looks like:

\begin{verbatim}
damp 0m0.amps;
damp 1m0.amps;
damp 1p0.amps;
damp 2m0.amps;
damp 2p0.amps;

realpar p0m0;
par p1m0;
par p1p0;
par p2m0;
par p2p0;

integral normInt(normInt.new);

event_loop:
    fcn = fcn - log(
        absSq(p0m0*0m0.amps + p1m0*1m0.amps
            + p1p0*1p0.amps + p2m0*2m0.amps
            + p2p0*2p0.amps)
    );

normalization:
    fcn = fcn + nevents * (
        p0m0*conj(p0m0)*normInt[0m0.amps , 0m0.amps] +
        p1m0*conj(p1m0)*normInt[1m0.amps , 1m0.amps] +
        p1p0*conj(p1p0)*normInt[1p0.amps , 1p0.amps] +
        p2m0*conj(p2m0)*normInt[2m0.amps , 2m0.amps] +
        p2p0*conj(p2p0)*normInt[2p0.amps , 2p0.amps] +

        2.0*real( p0m0*conj(p1m0)*normInt[0m0.amps , 1m0.amps] +
            p0m0*conj(p1p0)*normInt[0m0.amps , 1p0.amps] +
            p0m0*conj(p2m0)*normInt[0m0.amps , 2m0.amps] +
            p0m0*conj(p2p0)*normInt[0m0.amps , 2p0.amps]) +

        2.0*real( p1m0*conj(p0m0)*normInt[1m0.amps , 0m0.amps] +
            p1m0*conj(p1p0)*normInt[1m0.amps , 1p0.amps] +
            p1m0*conj(p2m0)*normInt[1m0.amps , 2m0.amps] +
            p1m0*conj(p2p0)*normInt[1m0.amps , 2p0.amps]) +

        2.0*real( p1p0*conj(p0m0)*normInt[1p0.amps , 0m0.amps] +
            p1p0*conj(p1m0)*normInt[1p0.amps , 1m0.amps] +
            p1p0*conj(p2m0)*normInt[1p0.amps , 2m0.amps] +
            p1p0*conj(p2p0)*normInt[1p0.amps , 2p0.amps]) +

        2.0*real( p2m0*conj(p0m0)*normInt[2m0.amps , 0m0.amps] +
            p2m0*conj(p1m0)*normInt[2m0.amps , 1m0.amps] +
            p2m0*conj(p1p0)*normInt[2m0.amps , 1p0.amps] +
            p2m0*conj(p2p0)*normInt[2m0.amps , 2p0.amps]) +

        2.0*real( p2p0*conj(p0m0)*normInt[2p0.amps , 0m0.amps] +
            p2p0*conj(p1m0)*normInt[2p0.amps , 1m0.amps] +
            p2p0*conj(p1p0)*normInt[2p0.amps , 1p0.amps] +
            p2p0*conj(p2m0)*normInt[2p0.amps , 2m0.amps]) 
    );
\end{verbatim}

This file shows all the key elements of the fit input file grammar.
The statements at the top of the file declare some variables.  {\tt
damp}'s are decay amplitudes, the string following is both the filename
where the amplitudes for a particular wave are found and the name of
the variable that can be used in the function to refer to this wave.
{\tt par}'s are the fit parameters, assumed to be complex unless {\tt
realpar} is specified.  Finally, {\tt integral normInt(normInt.new);}
declares {\tt normInt} to be a normalization integral found in the file
{\tt normInt.new}.  The executable statements in the second half of the
file are in two sections.  The {\tt event\_loop} section is executed for
every event in the dataset, while the {\tt normalization} section is
done only once, after the loop over events.  {\tt fcn} is a reserved
word that is the value to be minimized by varying the {\tt par}'s.
An astute reader might recognize {\tt fcn} as a relic of the {\sc
CERNLIB} {\tt minuit} minimizer, which is in fact the package used
to perform the actual minimization.  Notice also that the normalization
integrals are indexed by the name of the wave, or more precisely, the
name of the file that contained the decay amplitudes for that wave at
the time of integration.

The interpreter for this file is based heavily on {\tt hoc} by Kernighan
and Pike~\cite{book:kernighanandpike}.  Briefly, the file is parsed and statements are
stored as instructions to a virtual stack machine which is run at fit
time.  Decay amplitudes and parameters are stored in a symbol table and
their values are updated appropriately: every event for the amplitudes
and every iteration of the fitter for the parameters.  The code generated
for a part of the above event loop, {\tt p0m0*0m0.amps + p1m0*1m0.amps}
is shown in Table~\ref{fit_instructions}.
\begin{table}
\begin{center}
\begin{tabular}{|l|} \hline
varpush \\ \hline
{\tt p0m0} \\ \hline
eval \\ \hline
varpush \\ \hline
{\tt 0m0.amps} \\ \hline
eval \\ \hline
mul \\ \hline
varpush \\ \hline
{\tt p1m0} \\ \hline
eval \\ \hline
varpush \\ \hline
{\tt 1m0.amps} \\ \hline
eval \\ \hline
mul \\ \hline
add \\ \hline
\end{tabular}
\caption{\label{fit_instructions} Instructions generated by code fragment explained in text.}
\end{center}
\end{table}

Most instructions are simply pointers to a corresponding function,
therefore execution of the program means simply marching down this list
of function pointers, executing each one as you go.  Any ``instruction''
which is not a function pointer should be skipped by the true instruction
before it.  For instance, the initial varpush in the program shown
above pushes the next ``instruction'', the variable name {\tt p0m0}, onto the stack and increments
the program counter {\em  past } the next instruction to the eval that
follows it.  The eval function pops the top value off the stack, looks
it up in the symbol table, and pushes its {\em value} back onto the stack.
Therefore the first three instructions have the effect of pushing the
{\em value} of {\tt p0m0} onto the stack.  The next three instructions
similarly push the value of {\tt 0m0.amps} onto the stack.  The stack
now contains these two complex numbers, and the mul instruction pops both
off the stack, multiplies them, and pushes the result back on the stack.

This produces an extremely flexible fitting program, with two possible
drawbacks.  The first drawback is the complexity of the input file.  Even
the simple example given above generates many terms and a realistic fit
input using tens of waves could easily require pages of input containing
many similar symbols such as file names that differ by one character.
To reduce the opportunity for errors to arise, these input files are
often written by a separate program.  We use a {\sc prolog} program to
generate the {\tt fit} input files.  This program reads the states we
wish to include in the fit, produces a list of production amplitudes
to be used as fit parameters, and applies any known constraints such as
parity conservation to link any amplitudes it can.  The resulting list
is formatted appropriately for {\tt fit} to read and output to a file.

The second drawback of an interpreted system is performance.  While very
flexible, an interpreted system is inherently slower.  In recent work
where this has become an issue, the function written by the {\sc prolog}
program was modified to allow direct compilation by the C++ compiler
and was used directly with {\tt minuit} for fitting.

\subsection{Description of {\tt libpp} classes}
We have gathered into {\tt libpp} a collection of classes that were of
general use in particle physics, beyond the specialized topic of partial
wave analysis.  This section is not an attempt to fully document this
library, but rather just a sampling of a few of the objects available
to give the flavor of what is found in {\tt libpp.a}.
The complete documentation can be obtained from the source
using {\tt notangle}, or preformatted from the web at 
\verb+http://ignatz.phys.rpi.edu/~cummij/+

Three- and four-vectors are defined as {\tt threeVec} and {\tt
fourVec}, with most common operations available as member functions.
The implementation of {\tt fourVec}'s used contains a {\tt threeVec} and
a double, rather than deriving both from a base vector class.  While the
implementation details are not assumed in the interface to the class, one
can see reflections of the implementation in some of the methods such as
the constructors ({\tt fourVec(double t, threeVec space)}, for instance)

A particle data table class {\tt particleDataTable} is a simple
database containing information about known particle states from
the Particle Data Groups {\em Review of Particle Properties}.  It is
essentially a list of {\tt particleData} objects.  It is initialized
to a default set of values from the 2002 edition of the {\em Review
of Particle Properties}, but can be modified by reading a local file
containing ``custom'' particles.  Lookup in the table is implemented as
a linear search, which is not a performance issue for the typical size
tables involved.  The {\tt particle} class represents a physical instance
of a particle, {\em i.\ e.} a {\tt particleData} with an associated {\tt
fourVec}.  The {\tt event} class contains the beam, target, and a list
of final state particles.

Matrices are implemented as template classes to allow both the matrices of
complex numbers used in partial wave analysis and real matrices such as
Lorentz transformations to share the same code.  Lorentz transformations
are derived from the template class {\tt matrix<>} and can be used to
boost {\tt fourVec}'s, {\tt particle}'s or {\tt event}'s.  A particularly
useful constructor makes a {\tt lorentzTransform} from a {\tt fourVec},
defining the transformation to boost into the rest frame of the {\tt
fourVec}, treating it as a four-momentum.  This allows convenient
constructs such as:
\begin{verbatim}
        event e;
        // read the event from standard input
        std::cin >> e;
        lorentzTransform L(e.beam().get4P()+e.target().get4P());
        // put the event into the beam + target rest frame
        e = L*e;
\end{verbatim}
which puts the entire event into the center of mass system.

\section{Example analysis}
As a demonstration of the flexibility of this system, lets consider the
analysis of data from Jefferson Laboratory looking for ``missing baryon''
states.  The experiment collected data from the reaction $ \gamma p \to p
\pi^+ \pi^- $, with a $\gamma$ beam energy of 0.5-2.6 GeV/$c$.  Data were
selected which reconstructed all three final state particles, leaving a
data sample of 750k events for partial wave analysis.  While a complete
description of this analysis is beyond the scope of this paper, A brief
description of the analysis should illustrate the {\tt pwa2000} well.

This is a particularly difficult region to partial wave analyze, due
to the fact that the dynamics is expected to change drastically within
the range of the analysis.  For instance, for beam energies below $\rho$
threshold the data are in the resonance region and will probably be
well described by formation of isobars in the $s$-channel.  Above $\rho$
threshold the center of mass energy is leaving the resonance region and
the availability of the $\rho$ should result in a large contribution from
diffractive ($t$-channel) $\rho$ production.  The transition between these
two kinematic extremes will be difficult to map, and will require many
fits trying not only many different sets of waves but also different
likelihood functions as we make different assumptions about how to
describe $t$-channel processes in a truncated $s$-channel basis.  It is
for exactly this situation that we designed a flexible analysis system;
particularly, in this case, the interpreted fitting program: many different
function may be tried easily.

Initial fits, including only waves corresponding to isobar production
in the $s$-channel, were performed.  The production amplitudes obtained
from this fit are then used to calculate an acceptance corrected total
cross section.  The are plotted in Fig.~\ref{ppipi_sigma_tot}
\begin{figure}
\begin{center}
\includegraphics[width=4in]{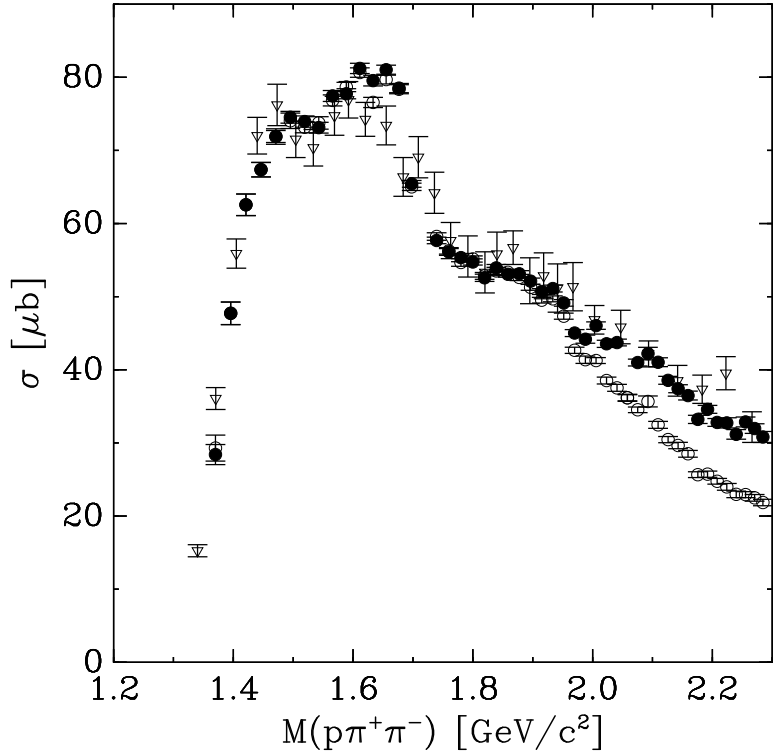}
\caption{\label{ppipi_sigma_tot}Results for the total cross section calculated from the partial wave analysis. Inverted triangles: Aachen-Berlin-Bonn-Hamburg-Heidelberg-M\"unchen collaboration.  Open circles: PWA using $s$-channel waves only.  Filled Circles: PWA using $s$-channel waves and $t$-channel $\rho$ production. }
\end{center}
\end{figure}
as open circles.  The points plotted as inverted triangles are from
an earlier bubble chamber experiment at DESY~\cite{ABBHHM:1968ke}.  The discrepancy above
1.9 GeV/c$^2$ is due to the fact that we are not using a sufficient
set of waves to describe our data.  It is easy to try including waves corresponding to
$t$-channel production of $\rho$'s: gamp generates the
amplitudes with a small change to the keyfile, and they can be included
into the fit either coherently or incoherently.  The result of such a
fit, with an incoherent $t$-channel $\rho$ production wave is plotted in
Fig.~\ref{ppipi_sigma_tot} as filled circles.  We can see the agreement
with the DESY experiment is much better, and the statistical errors,
even of this partial data set, are much smaller.

We can verify the fit describes the data better by using the fitted
production amplitudes and the calculated decay amplitudes to weight
events (generated uniformly in phase space) that passed through the
detector simulation.  This gives us a Monte-Carlo data set 
distributed in the kinematic variables as the fit found.  This Monte-Carlo data set can be
compared to the real data to measure the quality of the fit and give clues
about what waves need to be added.  For example, Fig.~\ref{fit_comp}
\begin{figure}
\begin{center}
\includegraphics[width=4in]{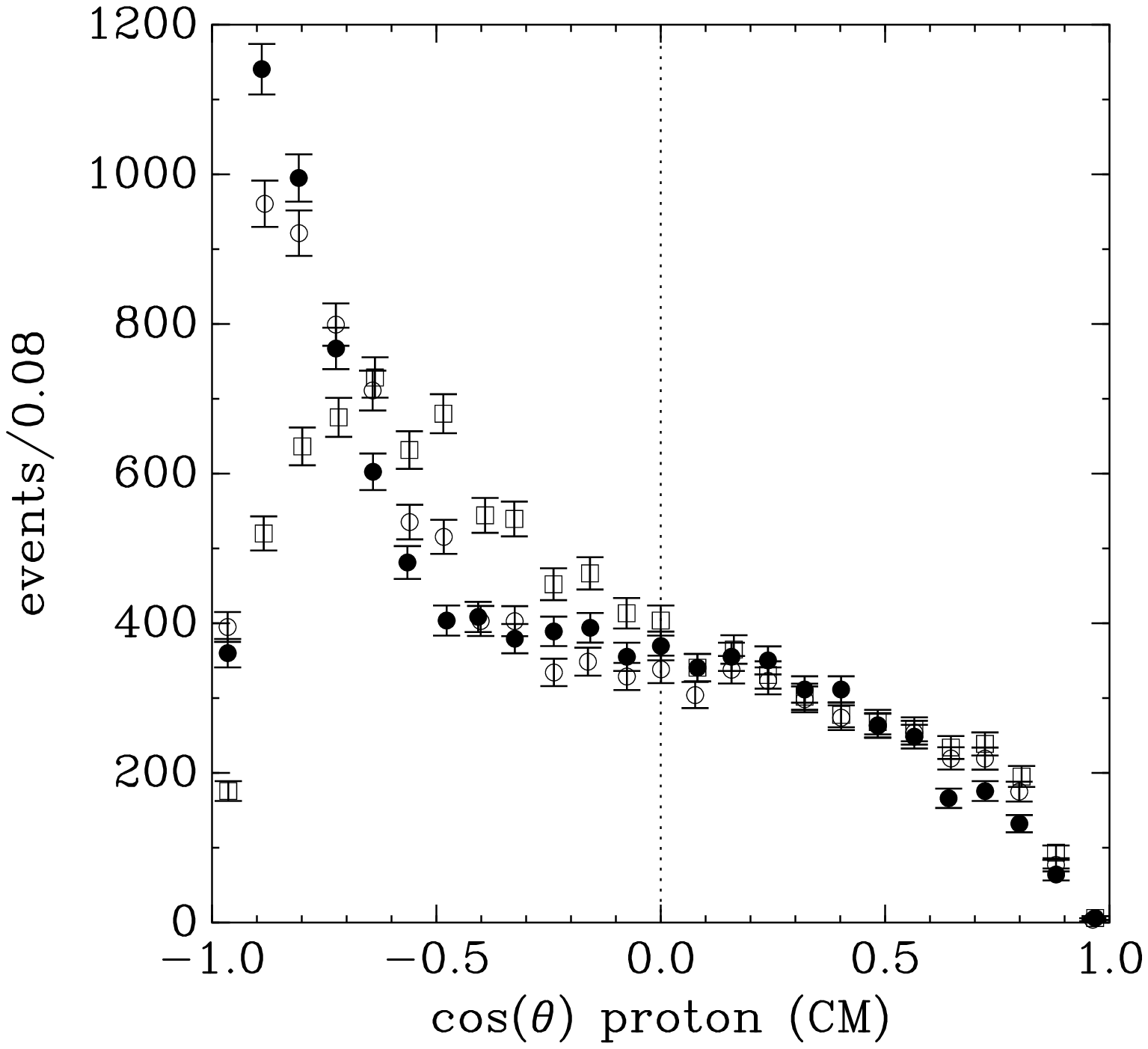}
\caption{\label{fit_comp} Comparison of $\cos(\theta_{cm})$ distributions of the proton for the data (solid circles), and the predictions of a fit using only $s$-channel waves (open squares), and a fit using $s$-channel waves with incoherent $t$-channel $\rho$ production (open circles).}
\end{center}
\end{figure}
shows the $\cos(\theta_{cm})$ distributions for the data (solid circles),
and the Monte-Carlo ``data'' sets generated from the fits mentioned above:
$s$-channel waves only and $s$-channel with incoherent $t$-channel $\rho$
production.  These distributions are for $2.06662 < M(p\pi^+\pi^-) <
2.08378$, where the total cross section is beginning to show discrepancy.
The fit using only $s$-channel waves (open squares) cannot create an
asymmetry large enough to describe the backward (in the center of mass
system) proton peak; however adding a simple, incoherent $t$-channel
$\rho$ production wave (open circle) improves the the description of
this kinematic variable quite a bit.

\section{Conclusions and Future Directions}

The {\tt pwa2000} is a flexible suite of tools developed for partial
wave analysis of particle physics data.  It has been used to analyze
peripheral meson production in a pion beam~\cite{Nozar:2002br}, peripheral meson
production in a photon beam, and $s$-channel baryon resonance
production in a photon beam~\cite{proc:bellisCIPANP}.  The object defined in the
library have been sufficient for all these analyses, the usual extent
of the modification necessary are to the input file parser to handle
input of information not required in previous analysis.  For instance,
the library implements the mass dependence of an isobar decay using an
abstract base class {\tt massDep}.  As additional parameterizations are
added for isobar decays, we add a derived class of {\tt massDep} that
implements the particular parameterization, add an appropriate keyword
to the keyfile language and modify the {\tt yacc} parser to instantiate
the new derived class when it sees the new keyword.

The strategy of separating the intelligence of the program from the
computation proved fruitful.  By using a separate program to generate
the input file for the fitting program, we were able to choose a language
well suited to each particular task.  {\sc prolog}, a language known for
its artificial intelligence applications, was used to apply the logical
constraints to the likelihood function such as physical conservation laws.
C++ routines were then driven by the input file {\sc prolog} writes.

%
Our experiences with the {\tt pwa2000} have shown us possible avenues for
future improvements.  The large statistics that will be coming available
from newer experiments will tremendously improve the statistical power
of the results.  Unfortunately this comes at a cost of speed: serial
fits for a single mass bin of projected data volumes could take days.
We have begun studies of parallelization of the maximum-likelihood fits.
One approach is to utilize work being done on Grid Computing~\cite{foster97globus}
or World-Wide Computing~\cite{varela:01a}.  Such an approach may be fruitful
for our problem since the evaluation of the likelihood function is
almost trivially parallelizable being a large sum over $\approx 10^6 -
10^8$ events.

The authors wish to thank all those who suffered through buggy versions of
this code.  Their patience and suggestions have produced a robust system.
This work was partially supported by the {\it National Science Foundation}.

\bibliography{refs}

\end{document}